\documentclass[a4paper, 11pt, conference]{ieeeconf}     
                                                        
\usepackage{graphics}
\usepackage{amsmath}
\usepackage{amssymb}
\usepackage{float}
\usepackage{listings} 
\usepackage{hyperref}
\usepackage[justification=centering]{caption}

\title{\LARGE \bf Georacle: Enabling Geospatially Aware Smart Contracts}
\author{Taha Azzaoui\\taha@azzaoui.org}

\begin{document}

\maketitle
\thispagestyle{empty}
\pagestyle{empty}

\begin{abstract}  
  Smart contracts \cite{Szabo_1997} have enabled a paradigm shift in
  computing by leveraging decentralized networks of trust to achieve consensus at scale.
  Oracle networks further extend the power of smart contracts by solving the
  so-called ``oracle problem'' \cite{breidenbach2021chainlink}. Such networks enable 
	smart contracts to make use of the vast amount pre-existing data available on the 
	web today without jeopardizing the integrity of the underlying network of trust. 
	By leveraging oracle networks, smart contracts can make decisions based on data
	corresponding to the physical world. To this end, we introduce \textbf{Georacle} \-- an 
	oracle service that enables geospatially aware smart contracts
  in a way that respects the space constrained nature of blockchain environments.
  Contracts can query the location of objects in a given
  area, map between street addresses and coordinates, and
  retrieve the geometry of a desired region of space while
  conserving gas consumption and avoiding unnecessary data processing.
\end{abstract}

\section{Introduction}
Geospatial data involves data about objects corresponding to a location on
the surface of the earth. Spatial information represented as vector data uses geometric
shapes such as points, lines, and polygons to represent geographic features in
space. This spatial information is usually combined with metadata containing
attribute information about the location allowing for efficient indexing and
retrieval. Once retrieved, geospatial data is often used for reasoning about
points of interest and determining relationships across regions in space.

In the current web landscape, location serves as a fundamental building block for
almost every useful application online today. From ride-sharing and vacation rentals
to social media and online dating, location has proven essential to fostering
digital ecosystems that rely on geospatial data to develop a computational understanding 
of the outside world.
In order for such ecosystems to thrive on the blockchain, decentralized applications
must have access to geospatial data on-chain. Native access to geodata empowers smart
contracts to coordinate across space in a similar fashion, forming the basis for
decentralized location-based experiences far more powerful than their centralized
counterparts on the web today. Using the transparency inherent in
blockchain-based ledger systems, contracts can make verifiable decisions on the
basis of location.

This paper introduces Georacle \-- an oracle service built on top of the 
Chainlink \cite{breidenbach2021chainlink} network with the goal of providing
smart contracts with the ability to query the attributes of arbitrary points in space.
Georacle serves as a map oracle, delivering location data to EVM-based 
blockchains \cite{wood2014ethereum} \cite{buterin2014next}. Among the contributions
of this paper is a novel approach for delivering
state of the art map data in a way that respects the space constrained
nature of blockchain environments. Contracts can query the location of
objects in a given area, map between street addresses and coordinates,
and retrieve the geometry of a desired region in space while conserving gas
utilization and avoiding unnecessary data processing.

\section{Data Model}
Georacle leverages OpenStreetMap \cite{OpenStreetMap} \-- a community-driven
open standard for interacting with geospatial data. The OpenStreetMap project
decentralizes the collection of geodata across thousands \cite{haklay2008openstreetmap}
of community members who serve as cartographers generating and validating geodata across
the world. As such, the OSM model replaces a central data authority with a community of voters
that discuss propositions to reach consensus on best practices. The project follows a peer-driven
approach similar to that of Wikipedia with the goal of providing open access to 
high quality geospatial data. At its core, the OSM model divides geodata into
an object hierarchy of increasing complexity as follows.

\subsection{Nodes}
Nodes are an atomic data type in the OpenStreetMap model, they encode a single
coordinate in space. A node is completely determined by its latitude and longitude.
Each node is assigned a unique identifier such that a contract can refer to a node
unambiguously using its ID.

\subsection{Ways}
Ways are the next level of abstraction in the OpenStreetMap data model, they
encode area features. Ways consist of a collection of nodes that 
combine to form some geometry. Closed ways (ones that start and end at the 
same node) are a special type of way called an area. Like nodes, contracts can 
also reference ways uniquely by their identifier.

\subsection{Tags}
Both nodes and ways can be tagged with metadata in the form of key-value pairs
that provide the object with some context (e.g. amenity=park, building=residential).
For a given object we might want to know if it has a name, what type of location it is, or what
its opening hours are. Tags provide contracts with the ability to filter locations
according to desired attributes. Figure \ref*{fig:etower} shows a selection of the available tags
for the way representing the Eiffel Tower.

\begin{figure}[H]
\begin{center}
\begin{tabular}{|c|c|}
\hline
Key               & Value                 \\ \hline
addr:city         & Paris                 \\ \hline
addr:housenumber  & 5                     \\ \hline
addr:postcode     & 75007                 \\ \hline
addr:street       & Avenue Anatole France \\ \hline
architect         & Stephen Sauvestre     \\ \hline
building          & attraction            \\ \hline
building:colour   & \#706550              \\ \hline
building:material & iron                  \\ \hline
building:shape    & pyramidal             \\ \hline
fee               & 10-25€                \\ \hline
height            & 324                   \\ \hline
\end{tabular}
\end{center}
  \caption{A Selection of Tags For the Eiffel Tower (Way 5013364)}
  \label{fig:etower}
\end{figure}

\section{Location Queries}
By combining location and metadata, contracts can filter objects by their tags
to query regions of interest with desired attributes across space. This direct access to
the OSM data model offers contracts a general method for constructing spatially
constrained queries with desired levels of complexity.
Contracts can search globally for a list of points tagged with some key-value pair or opt for a
more local search by specifying an arbitrary bounding box around the region of
interest. On the other hand, if the points of interest are known a priori, contracts can
filter the latest metadata associated with the specific object to learn more about that region of space.

Armed with these techniques, smart contracts are not only geospatially aware,
but can also be made contextually aware with respect to the locations they interact
with. That is, metadata allows contracts to learn more about the nature of a place
(e.g. is it a building, shop, park, ATM, etc) along with location dependent
information such as the opening hours of a given building or any fees associated
with entering. Using this information, contracts can make decisions with
implications that reach into the physical world by conditioning their logic
on the attributes associated with their points of interest. Use cases such as
crowd funding and decentralized autonomous organizations can function on the
basis of location by tailoring queries to fit their needs. 

\subsection{Area}
One method of informing contracts about the outside world involves querying
nodes or ways confined to a named area that are tagged with a desired description.
The \textit{nodesInArea} function and its corresponding way variant \textit{waysInArea}
both expect a named area (e.g. New York, London, Tokyo, etc) along
with a key-value pair of desired attributes and an upper bound on the number of
search results. Figure \ref*{fig:nodesInArea} shows an example of searching for $n$ coffee
shops (represented as nodes) within the Boston area by crafting a Chainlink request for
Georacle using Solidity \cite{solidity}. Upon fulfillment, the oracle will respond with
an \textbf{int64} array of size at most $n$ representing the matching object identifiers packed
according to the EVM ABI specification.

\begin{figure}[H]
\begin{lstlisting}
req.add("function","nodesInArea");
req.add("key","amenity");
req.add("value","cafe");
req.add("area","Boston");
req.addInt("limit", n);
\end{lstlisting}
\caption{Finding $n$ Coffee Shops in the Boston Area}
\label{fig:nodesInArea}
\end{figure}

Additionally, contracts can obtain a count of the number
of matching identifiers beforehand by using the \textit{nodeCountInArea}
and \textit{wayCountInArea} functions as shown in Figure
\ref*{fig:nodeCountInArea}. Upon fulfillment, the oracle will return a
single \textbf{int64} value denoting the number of objects tagged with the
requested description in the area.

\begin{figure}[H]
\begin{lstlisting}
req.add("function", "nodeCountInArea");
req.add("key", "amenity");
req.add("value", "cafe");
req.add("area", "Boston");
\end{lstlisting}
\caption{Retrieving a Count of Coffee Shops in the Boston Area}
\label{fig:nodeCountInArea}
\end{figure}

\subsection{Bounding Box}
When interacting with less well-defined areas, it can be useful to work with a custom
bounding box surrounding some region of concern rather than a named area.
Using a bounding box allows contracts to confine their queries
to an arbitrary region of space. This is useful if the desired search space
crosses the boundaries of multiple named areas. For regions that are known to be some
subregion of a named area, using a bounding box will return only the matches within
this desired subspace, incurring less data transfer and thus consuming less gas.

Figure \ref*{fig:nodesInBB} searches for $n$ train stations within a subregion of Manhattan
using the bounding box: $(40.7719, -73.9746, 40.7975, -73.9469)$. These
coordinates correspond to the south, west, north, and east most points of
the bounding box respectively. Note that Georacle scales coordinates by a
factor of $10^8$ as the EVM lacks support for floating point arithmetic.
Upon fulfillment, the oracle response format will be identical to that of the named
area queries introduced previously.

\begin{figure}[H]
\begin{lstlisting}
req.add("function", "nodesInBB");
req.add("key", "public_transport");
req.add("value", "station");
req.addInt("south", 4077190000);
req.addInt("west", -7397460000);
req.addInt("north", 4079750000);
req.addInt("east", -7394690000);
req.addInt("limit", n);
\end{lstlisting}
\caption{Obtaining $n$ Train Stations in a Subregion of Manhattan}
\label{fig:nodesInBB}
\end{figure}

Like with named areas, contracts can also retrieve a count of the number
of matching identifiers within the bounding box beforehand by using the \textit{nodeCountInBB} and
\textit{wayCountInBB} functions as shown in Figure \ref*{fig:nodeCountInBB}.
Upon fulfillment, the oracle will return a single \textbf{int64} value
denoting the number of objects tagged with the requested description within the
specified bounding box.

\begin{figure}[H]
\begin{lstlisting}
req.add("function", "nodeCountInBB");
req.add("key", "public_transport");
req.add("value", "station");
req.addInt("south", 4077190000);
req.addInt("west", -7397460000);
req.addInt("north", 4079750000);
req.addInt("east", -7394690000);
\end{lstlisting}
\caption{Obtaining the Number of Train Stations Within a Subregion of Manhattan}
\label{fig:nodeCountInBB}
\end{figure}

\subsection{Filtering Tags}
If the regions of interest are known a priori, either because they are embedded in the smart
contract itself, or because they are the return value of one of the functions introduced
previously, the function \textit{nodeTagQuery} (along with its way variant
\textit{wayTagQuery}) provides contracts with a mechanism for filtering the tags of a
specific object. Both functions expect a string array of keys along with the known object
identifier.

\begin{figure}[H]
\begin{lstlisting}
string[] memory tags = [
  "name",
  "addr:housenumber",
  "addr:street",
  "addr:city",
  "addr:postcode",
  "opening_hours"];
req.add("function", "nodeTagQuery");
req.addInt("ID", 2700809522);
req.addStringArray("tags", tags);
\end{lstlisting}
\caption{Filtering the Tags of a Local Coffee Shop (Node 2700809522)}
\label{fig:nodeTagQuery}
\end{figure}

Figure \ref*{fig:nodeTagQuery} shows the process of retrieving the name,
address and opening hours of a specific coffee shop in the Boston area.
Upon fulfillment, the oracle will respond with an ABI-packed \textbf{string}
array of values corresponding to each key.

\subsection{Geometry}
In addition to querying discrete points, contracts can also interact with the physical
geometry of a given region. Recall that ways are simply a collection of nodes
(points in space). The \textit{wayGeometry} function can be used to obtain the
coordinates of the underlying collection of nodes that represent a given way. These 
nodes form some geometry which can then be used to compute geometric properties like
area or input into common computational geometric algorithms such as nearest neighbor
or segment intersection. Figure \ref*{fig:wayGeometry} shows the retrieval of the way geometry
representing the Great Pyramid of Giza. This geometry consists of
$10$ nodes which combine to form the rectangular base of the pyramid.

\begin{figure}[H]
\begin{lstlisting}
req.add("function", "wayGeometry");
req.addInt("ID", 4420397);
\end{lstlisting}
\caption{Obtaining the Geometry of the Great Pyramid of Giza (way 4420397)}
\label{fig:wayGeometry}
\end{figure}

Upon fulfillment, the oracle will return a list of pairs of \textbf{int64}
values representing the latitude and longitude coordinates of each node.
Due to the topology of large areas however, geometric data can sometimes be
prohibitively expensive to store completely on chain. To gauge gas consumption, it
can be useful to obtain the size of the geometry beforehand by using the \textit{wayCount}
function as shown in figure \ref*{fig:wayCount}. Upon fulfillment, the oracle will return 
a single \textbf{int64} value representing the number of nodes that form the way geometry.

\begin{figure}[H]
\begin{lstlisting}
req.add("function", "wayCount");
req.addInt("ID", 4420397);
\end{lstlisting}
\caption{Retrieving the Number of Nodes in the Geometry of the Great Pyramid of Giza (way 4420397)}
\label{fig:wayCount}
\end{figure}

\section{Geocoding}
While coordinates lend themselves well to geometric reasoning, it can be
convenient for users to refer to a location based on its canonical street addresses.
Smart contracts can use the \textit{geocode} function to map an area description
(i.e. street address) to the coordinates of its representative OSM object.
Figure \ref*{fig:geocode} obtains the coordinates of a street address in the United Kingdom.

\begin{figure}[H]
\begin{lstlisting}
req.add("function", "geocode");
req.add("address",
        "221B Baker St, London NW1 6XE, UK");
\end{lstlisting}
\caption{Mapping a Street Address to a Point in Space}
\label{fig:geocode}
\end{figure}

Upon fulfillment, the oracle will respond with an ABI-packed \textbf{struct} of four
\textbf{int64} values representing an object type flag ($0 =$ node, $1 =$ way), 
the object identifier, and the corresponding latitude and longitude of the area respectively.
This output can then be used to preform spatial analysis on user-provided location
information, which is most commonly in the form of a description.

Smart contracts can also use the \textit{reverseGeocode} function to obtain the
inverse mapping from a set of coordinates to a description of the nearest OSM object
as shown in figure \ref*{fig:reverseGeocode}.

\begin{figure}[H]
\begin{lstlisting}
req.add("function", "reverseGeocode");
req.addInt("lat", 5152338790);
req.addInt("lon", -15823670);
\end{lstlisting}
\caption{Mapping a Coordinate Pair to an Area Description}
\label{fig:reverseGeocode}
\end{figure}

Upon fulfillment the oracle will respond with an ABI-packed \textbf{struct} consisting of
two \textbf{int64} values representing an object type flag and the corresponding object
identifier along with a \textbf{string} description of the object.

\section{Gas Considerations}
As the demand for block space increases, optimizing gas consumption is of primary concern for
the practical usage of decentralized applications. The OSM data model outlined in the previous
sections is well-suited for blockchain environments where data storage is at a premium, since
contracts need only interact with objects that fit their desired description. Application
developers can assess gas consumption beforehand by simulating oracle queries off-chain and
determining the subset of data necessary to manipulate on chain. Contracts can then optimize for
gas by fine-tuning their search space and imposing an upper bound on the number of returned
search results. This upper bound can be computed dynamically based on the count
variant associated with each query function.

\section{Conclusion}
Building location aware smart contracts involves bringing geodata
on-chain in a manner that respects the space constrained nature of modern
blockchains. The OSM data model and its hierarchy of uniquely identified
object types is well-suited for this task as location representation reduces to
the corresponding object identifier.
The metadata associated with each object allows for flexible queries that can be
confined to a named area or fine-tuned to a specific region in space on the fly.

While named areas can be a convenience for users, the ability to specify an arbitrary
search space allows smart contract developers to narrow down the regions that
matter most and avoid filtering through unnecessary data on chain. Geocoding gives 
smart contracts the ability to translate between sets of coordinates and human
readable area descriptions better suited for interfacing with users.
Smart contracts that leverage geospatial data by combining these mechanisms can coordinate
across space, creating an ecosystem with global implications. By querying the attributes
of specific points of interest, contracts are no longer insulated from the outside world
as they gain a better sense of their surroundings.

\bibliographystyle{plain}
\bibliography{whitepaper}

\begin{thebibliography}{1}

\bibitem{breidenbach2021chainlink}
Lorenz Breidenbach, Christian Cachin, Benedict Chan, Alex Coventry, Steve
  Ellis, Ari Juels, Farinaz Koushanfar, Andrew Miller, Brendan Magauran, Daniel
  Moroz, et~al.
\newblock Chainlink 2.0: Next steps in the evolution of decentralized oracle
  networks, 2021.

\bibitem{buterin2014next}
Vitalik Buterin et~al.
\newblock A next-generation smart contract and decentralized application
  platform.

\bibitem{solidity}
{Christian Reitwiessner and Gavin Wood}.
\newblock \url{https://docs.soliditylang.org/}, 2015.

\bibitem{haklay2008openstreetmap}
Mordechai Haklay and Patrick Weber.
\newblock Openstreetmap: User-generated street maps.
\newblock {\em IEEE Pervasive computing}, 7(4):12--18, 2008.

\bibitem{OpenStreetMap}
{OpenStreetMap contributors}.
\newblock {Planet dump retrieved from https://planet.osm.org }.
\newblock \url{https://www.openstreetmap.org}, 2017.

\bibitem{Szabo_1997}
Nick Szabo.
\newblock Formalizing and securing relationships on public networks.
\newblock {\em First Monday}, 2(9), Sep. 1997.

\bibitem{wood2014ethereum}
Gavin Wood.
\newblock Ethereum: A secure decentralised generalised transaction ledger.

\end{thebibliography}
\end{document}